\documentstyle[12pt]{article}

\begin{document}

\begin{center}

{\Large \bf {Dilaton Field and Massless Particle

 for 2d Gravity}}

\vspace{3mm}

{\bf {George Chechelashvili and George Jorjadze}}

\vspace{3mm}

{\it {Dept. of Theoretical Physics, Razmadze Mathematical
Institute\\
Tbilisi, Georgia\\}}

\end{center}

\begin{abstract}
\noindent
It is shown that the model of 2d dilaton gravity is equivalent
to the dynamical system of massless particles in the Liouville field.
\end{abstract}

\section{Introduction}

The  Einstein theory of gravity in 2-dimensions is degenerated [1].
The corresponding action
\begin{equation}
S_0 =\int d^2x~\sqrt{-g}~ R
\end{equation}
is expressed only through the `surface terms' and its variation
identically vanishes.
As a result, the Einstein equations $R_{ab}-1/2~ R g_{ab}=0$ in 2-dimensions are
identities rather than dynamical equations.

The covariant equations $R_{ab}-1/2~ \Lambda g_{ab}=0$, where $\Lambda$ is
a constant, can be considered as a model of 2d general relativity.
These equations are equivalent to $R=\Lambda$, but
construction of the corresponding covariant action only in terms of the
metric tensor $g_{ab}$ is problematic.

Locally, any 2d Lorentzian manifold is conformally flat [2], i.e.
in some local coordinates the metric tensor can be written in the form
\begin{equation}
 g_{ab}(X) = \exp{q(X)}\;\;\left( \begin{array}{cr}
 1 & 0 \\ 0 & -1
 \end{array} \right),
 \end{equation}
where $X:=(x^0,x^1)$ and $q(X)$ is a field.
The constant curvature manifold in the conformal gauge
(1.2) is described by the  Liouville equation [3]
\begin{equation}
(\partial ^2_0 - \partial ^2_1)\;q(X)  + \Lambda e^{q(X)} = 0,
\end{equation}
which is a variational equation for the functional
\begin{equation}
S[q] =1/2\int d^2x~\left [(\partial_0q)^2 -(\partial_1q)^2
-2\Lambda e^q\right ].
\end{equation}
Action (1.4) has non-covariant form from the point of view of general
relativity  and to obtain the corresponding covariant action
it is necessary to introduce some additional auxiliary fields.
A covariant model with minimal extension is 2d dilaton gravity [4],
which arises in Polyakov string theory [5].

For a system with reparametrization invariant
action and auxiliary fields, usually,
it is difficult to guess what are the physical variables
and what kind of physical degrees of freedom they define.
Such analysis can be done using the Dirac's procedure for
constrained systems based on the Hamiltonian reduction [6].
In the case  of 2d dilaton gravity,
this procedure eliminates all field degrees of freedom
and after full Hamiltonian reduction the system only
with finite degrees of freedom remains [7].

In this paper we compare the model of
2d dilaton gravity with the system of massless
particle in the corresponding Liouville field [8] and demonstrate
their equivalence.

The paper is organized as follows: in Section 2 we prove degeneracy
of 2d Einstein gravity. Obtained formulas are used
in Section 3 for the description of 2d dilaton gravity.
In Section 4 we consider the examples of constant curvature manifolds
with global symmetries and analyze
the corresponding models of 2d dilaton gravity.
In Conclusions we show equivalence of the considered models of 2d gravity
and the dynamical systems of massless particles.

\setcounter{equation}{0}

\section{Degeneracy of general relativity in 2-dimensions}

The Einstein-Hilbert Lagrangian
${\cal L}_0 :=\sqrt{-g}~ R$ can be splitted in two parts [1]
\begin{equation}
{\cal L}_0={\cal L}_1 +{\cal  L}_2,
\end{equation}
where the first term ${\cal L}_1$ contains derivatives of the metric tensor
$g_{ab}$ only in the first order
\begin{equation}
{\cal L}_1 =\frac{1}{2}\Gamma_{ab}^b\partial_c(\sqrt{-g}~g^{ac}) -
\frac{1}{2}\Gamma_{ab}^c\partial_c(\sqrt{-g}~g^{ab}),
\end{equation}
with
\begin{equation}
\Gamma_{ab}^c :=\frac{1}{2}g^{cd}(\partial_a g_{bd}+\partial_b g_{ad}-
\partial_d g_{ab})
\end {equation}
and the second term ${\cal L}_2$ is a `total derivative'
\begin{equation}
{\cal L}_2 =\partial_c\left [\sqrt{-g}~(g^{ab}\Gamma_{ab}^c -
g^{ac}\Gamma_{ab}^b)\right ].
\end{equation}
As a result, ${\cal L}_2$ has no contribution in the variational
equations
$R_{ab}-1/2~ Rg_{ab} =0$
and the Einstein tensor $G_{ab}:=R_{ab}-1/2~Rg_{ab}$ is defined only
through ${\cal L}_1$
\begin{equation}
G_{ab} =\frac{1}{\sqrt{-g}}\left (\frac{\partial {\cal L}_1 }{
\partial g^{ab}} -\partial_c \frac{\partial {\cal L}_1 }{
\partial~\partial_c g^{ab}}\right ).
\end{equation}
Eqs. (2.1)-(2.5) are valid for arbitrary dimension ($D\geq 2$) of spacetime,
but the case $D=2$ posses certain singularity.
The Reimann tensor of 2d manifold has only one independent component
$R_{0101}$ and the corresponding Einstein tensor $G_{ab}$
identically vanishes for arbitrary metric tensor $g_{ab}$.
Therefore, 2d Einstein equations are identities
and not the dynamical equations for $g_{ab}$. This degeneracy can be
seen from the Einstein-Hilbert action (1.1) as well.

Writing the metric tensor of 2d spacetime in the form
\begin{equation}
g_{ab}= \left( \begin{array}{cr}
 A & B \\ B & C
 \end{array} \right)
\end{equation}
and using (2.2)-(2.4) we get
\begin{equation}
{\cal {L}}_1 =-\frac{1}{2G^{3}}[A(\dot B C^\prime - \dot CB^\prime )
+B(\dot C A^\prime - \dot A C^\prime )
+C(\dot A B^\prime - \dot B A^\prime )],
\end{equation}
\begin{equation}
{\cal L}_2 =\partial_t\left (\frac{\dot C-B^\prime}{{G}}\right )
+\partial_x\left (\frac{A^\prime -\dot B}{{G}}\right ),
\end{equation}
where
\begin{equation}
G:=\sqrt{-g} = \sqrt{B^2-AC}
\end{equation}
and we also use the notations
$x^0:=t,~ x^1:=x;~ \partial_t A:=\dot A,~ \partial_x A:= A^\prime,...$.

Note that (2.7) can be considered as a value of the 2-form
\begin{equation}
\omega =-\frac{1}{2G^{3}}(A~dB\wedge dC + B~dC\wedge dA
+C~dA\wedge dB),
\end{equation}
on the tangent vectors $V_t:= (\dot A,\dot B, \dot C)$ and
$V_x :=(A^\prime , B^\prime , C^\prime  )$:
\begin{equation}
{\cal {L}}_1 =
\omega (V_t, V_x)
\end{equation}
Since 2-form (2.10) is exact $\omega =d\theta$ [9],
(2.11) can be written as
\begin{equation}
{\cal {L}}_1 = \partial_t \theta (V_x) -\partial_x \theta (V_t).
\end{equation}
Thus, in 2-dimensions, ${\cal L}_1$ is also a total derivative.

The 1-form $\theta$ we can choose by
\begin{equation}
\theta =\frac{1}{{G}}\left (\frac{B}{C}dC - dB\right )
\end{equation}
and finally, from (2.1), (2.8) and (2.12), we get
\begin{equation}
\sqrt{-g}R(g_{ab}) =\partial_t\left[\frac{1}{G}
\left (\dot C-2B^\prime +B\frac{C^\prime}{C}\right )\right ]
+\partial_x\left[\frac{1}{{G}}
\left (A^\prime -B\frac{\dot C}{C}\right )\right ].
\end{equation}

\setcounter{equation}{0}
\section{The model of 2d gravity with dilaton field}

Let us consider a model of 2d gravity defined by the action [4]
\begin{equation}
S=\int d^2x~ \sqrt{-g}~\left [
 \frac{1}{2}g^{ab}\partial_a\phi\partial_b\phi
-\phi R +\Lambda  \right ],
\end {equation}
where
$\phi$ is a scalar (dilaton) filed and
$\Lambda$ is a `cosmological' constant.

Variation of (3.1) with respect to the dilaton field defines the
dynamical equation
\begin{equation}
-\frac{1}{\sqrt{-g}}\frac{\delta S}{\delta \phi} =
\nabla^2\phi +R = 0.
\end {equation}

To calculate variations of (3.1) with respect to the gravitational
field $g_{ab}$ we use the analysis of the Einstein-Hilbert Lagrangian
${\cal L}_0 :=\sqrt{-g}~R(g_{ab})$ given in the previous section.
Using (2.1)-(2.4) and neglecting the boundary terms, the functional
$S_d := \int d^2x~  \sqrt{-g}~\phi R $
can be written as
\begin{equation}
S_d \simeq \int d^2x (\phi {\cal L}_1 -\partial_a \phi X^a ),
~~~\mbox{with}~~~ X^a = \sqrt{-g}~(g^{bc}\Gamma_{bc}^a -
g^{ab}\Gamma_{bc}^c).
\end{equation}
The variation of (3.3) reads
\begin{equation}
\frac{1}{\sqrt{-g}}\frac{\delta S_d}{\delta g^{ab}} = \phi G_{ab}
+(g_{ab} \nabla^2 -\nabla_a\nabla_b)\phi ,
\end {equation}
where $\nabla_a\nabla_b\phi = \partial^2_{ab}\phi +\Gamma^c_{ab}\partial_c
\phi$.

Eq. (3.4) valid for arbitrary dimension of spacetime.
Since in 2-dimensions the Einstein tensor $G_{ab}$ vanishes, for
the variational equations of (3.1) we obtain
\begin{equation}
\frac{2}{\sqrt{-g}}\frac{\delta S}{\delta g^{ab}} =
\partial_a\phi\partial_b\phi + g_{ab}\left
(-\frac{1}{2}g^{cd}\partial_c\phi\partial_d\phi -\Lambda\right )
-2(g_{ab} \nabla^2 -\nabla_a\nabla_b)\phi = 0.
\end {equation}
Note that eq.(3.4) in 2-dimensions can be verified from (2.14) as well.

Taking trace of (3.5) and using (3.2) we get
that the dynamical system (3.1) describes a spacetime manifold
with a constant curvature $R = \Lambda$.

The conformal gauge (1.2) leads to the following form of the
dynamical equations (3.2) and (3.5)
\begin{equation}
\partial_{+,-}^2 \varphi =0,~~~~~~~~~~
4\partial_{+,-}^2 q = {\Lambda}e^q.
\end{equation}
\begin{equation}
\left (\partial_+\varphi \right)^2 -2\partial^2_{+,+}\varphi
=\left (\partial_+ q \right)^2 -2\partial^2_{+,+} q,~~~~~~~
\left (\partial_-\varphi \right)^2 -2\partial^2_{-,-}\varphi
=\left (\partial_- q \right)^2 -2\partial^2_{-,-} q,
\end{equation}
where $x^\pm :=x^1\pm x^0$ are the
light cone coordinates and $ \varphi := q -\phi$.

As it was expected,
the field $q(X)$ satisfies the Liouville equation (1.3), while
$\varphi (X)$ is a `free' field. According to (3.7),
the traceless energy-momentum
tensors [10-11] of these two fields are equal to each other.

The general solution of the Liouville equation has the form [3]
\begin{equation}
q(x^+,x^-) = \log \frac{8A^{+~\prime}(x^+)A^{-~\prime}(x^-)}{|\Lambda |
\left [A^+(x^+)+\epsilon A^-(x^-)\right ]^2}
\end{equation}
where $\epsilon$
is a sign of $\Lambda$ ($\epsilon =\Lambda/|\Lambda|$) and
$A^\pm$ are any monotonic functions with $A^{\pm~\prime}>0$.

The conformal form of the metric tensor (1.2) fixes the gauge freedom
only up to the conformal transformations
\begin{equation}
x^\pm \mapsto \tilde x^\pm ;~~~x^\pm =f^\pm(\tilde x^\pm)
\end{equation}
The corresponding transformation of the field $q$ is given by
\begin{equation}
q(x^+,x^-) \mapsto \tilde q(x^+,x^-)=q (f^+(x^+), f^-(x^-) )
+\log f^{+~\prime} (x^+) f^{-~\prime} (x^-).
\end{equation}

The general solution (3.8) can be
obtained by the conformal transformation (with $f^\pm(x^\pm)=A^\pm (x^\pm)$)
of the Liouville field
\begin{equation}
q(x^+,x^-) = \log \frac{8}{|\Lambda |
(x^++\epsilon x^-)^2}.
\end{equation}
Hence, all Liouville fields (with fixed $\Lambda$) are related
to each other by the conformal transformations (3.9)-(3.10).
Since the considered system is reparametrization invariant,
one can fix the field $q(X)$
by suitable choice of local coordinates.

After fixing the Liouville field $q(X)$
the dynamical freedom of the system is described by the free field
$\varphi (X)$, which should be defined
from (3.7).  These equations are
ordinary differential equations for two chiral parts
$\varphi_\pm (x^\pm)$ of the free field
$\varphi (X)= \varphi_+ (x^+) +\varphi_- (x^-)$.
For a fixed right hand side of (3.7)
the corresponding freedom is described by a finite number of
integration constants.

Thus, the system (3.1) has no physical degrees of freedom for
field variables and full gage fixing leads
to a finite dimensional system.
This result can be obtained by the Hamiltonian reduction procedure
as well (see [7]).

It should be noted that
the conjecture about the conformal form of
the metric tensor (1.2) is valid only locally.
Therefore, constructions of this section have only a local character.
To get the global picture of the system
one should specify the global properties of spacetime.

\setcounter{equation}{0}

\section{The examples of 2d dilaton gravity}

The one-sheet hyberboloid
\begin{equation}
-(y_0)^2+(y_1)^2+(y_2)^2=m^{-2},
\end{equation}
embedded in 3-dimensional Minkowski space  is
the example of Lorentzian manifold with constant curvature  [2].
The parameter $m$ ($m>0$) defines the scalar curvature  $R=-2m^2$.
It is obvious that the isometry group of the hyperboloid is
$SO_\uparrow (2.1)$.

Using the parametrization
\begin{eqnarray}
y_0=-\frac{\cot m\rho}{m},~~~~y_1=\frac{\cos m\theta}{m\sin
m\rho},~~~~y_2=\frac{\sin m\theta}{m\sin m\rho},\nonumber \\
{\mbox {where}}~~~~ \rho \in ]0,\pi/m[,~~~\theta \in [0,2\pi/m[
~~~~~~~~~~~~~~~~~~~~~
\end{eqnarray}
we get the conformal form (1.2) of the induced metric tensor
\begin{equation}
g_{ab}(\rho,\theta)=\frac{1}{\sin^2 m\rho}
\left( \begin{array}{cr}
1&0\\0&-1 \end{array} \right),
\end{equation}
which defines the Liouville field (1.3)
\begin{equation}
q(\rho,\theta)=-\log \sin^2 m\rho ,
\end{equation}
for $\Lambda =-2m^2$.
The parameters $\rho$ and $\theta$ can be identified with
the spacetime coordinates $x^0$ and $x^1$, respectively. The
corresponding light-cone coordinates we denote by $\theta_\pm$
($\theta_\pm :=\theta\pm\rho $).

The traceless energy momentum tensor for (4.4) is given by
\begin{equation}
T_{++}:=
\left (\partial_+ q \right)^2 -2\partial^2_{+,+} q =-m^2,~~~~~~~
T_{--}:=\left (\partial_- q \right)^2 -2\partial^2_{-,-} q=-m^2.
\end{equation}
and (3.7) defines the following
two equations for the chiral $\varphi_+(x^+)$
and anti-chiral $\varphi_-(x^-)$ parts of the free field  $\varphi$
\begin{equation}
(\varphi_+')^2 -2(\varphi_+'') =-m^2,~~~~~~~
(\varphi_-')^2 -2(\varphi_-'') =-m^2.
\end{equation}
After integration we get
\begin{equation}
\varphi =\varphi_+ +\varphi_- =C -\log \sin^2 \frac{m}{2}(\theta_+-\alpha_+)-
\log \sin^2 \frac{m}{2}(\theta_- -\alpha_-),
\end{equation}
where $\alpha_\pm$ and $C$ ($0\leq \alpha_\pm<2\pi/m$)
are integration constants.
Actually, these three parameters describe  all freedom of the system.

The general solution (4.7) has two lines of singularity at
$\theta =\alpha_\pm \mp\rho$.
According to (4.2), the coordinates of
singularity lines on the hyperboloid (4.1) satisfy the relations
\begin{equation}
y_0+y_1\cos m\alpha_\pm+y_2\sin m\alpha_\pm=0,~~~~~~y_1\sin m\alpha_\pm
-y_2\cos m\alpha_\pm =\mp\frac{1}{m},
\end{equation}
which define generatrices of the hyperboloid.

The Cauchy data for the field $\varphi (\rho ,\theta)$ at the moment
$\rho =\rho_0$ is given by two functions
\begin{equation}
\Phi_{\rho_0}(\theta ):=\varphi (\rho_0 ,\theta )~~~~~
\mbox {and}~~~~
\Pi_{\rho_0}(\theta ):=\partial_\rho\varphi (\rho_0 ,\theta ).
\end{equation}
These functions are singular at the points $\theta =\alpha_\pm \mp \rho_0$.
When these points are different
($\alpha_+ - \rho_0 \neq \alpha_-+ \rho_0$),
behaviour of the functions (4.9) near the singular points
is given by (see (4.7))
\begin{equation}
\Phi_{\rho_0}(\theta )\sim\log\frac{4}{m^2(\theta -\alpha_\pm\pm\rho_0)^2},
~~~~~~~~~
\Pi_{\rho_0}(\theta )\sim\mp\frac{2}{\theta -\alpha_\pm\pm\rho_0}.
\end{equation}
If $\rho_0$ is a `moment' of intersection of singularity lines,
the singularity points coincide
($\alpha_{+}-\rho_0 = \alpha_-+ \rho_0 :=\theta_0$ ).
In that case $\Pi_{\rho_0}(\theta )$ is regular and
$$
\phi_{\rho_0}(\theta )\sim\log\frac{8}{m^2(\theta -\theta_0)^2}.
$$
Thus, the character of singularities of the Cauchy data
is preserved in dynamics
if the singularity lines does not intersect each other.
This case correspond to $\alpha_+=\alpha_-$.

Another example of spacetime manifold with constant negative curvature
can be related to the Liouville field (3.11) for $\Lambda < 0$.
This field is singular at $x^0 =0$. If one excludes the singularity
line (see Fig.1), the domains $x^0>0$
and $x^0<0$ can be considered as two different patches of the
hyperboloid (4.1). The corresponding conformal map to the
coordinates $(\rho, \theta)$ (see (4.2)) can be written as
\begin{equation}
x^\pm = \frac{1}{m}\tan \frac{m}{2}(\theta \pm \rho ).
\end{equation}
%This map covers almost all hyperboloid, except two generatrices.

The traceless energy momentum tensor for the Liouville
field (3.11) identically vanishes
and (3.7) defines the following free field
\begin{equation}
\varphi =  -2\log M^2|(x^+-\beta_+)(x^--\beta_-)|
\end{equation}
where $\beta_\pm$ and $M$ are the integration constants.
The free fields (4.7) and (4.12) are related by the conformal map
(4.11) and the case with non intersecting singularities
($\alpha_+=\alpha_-$) correspond to  $\beta_+=\beta_-$
(see Fig.1).

\vspace{0.5cm}

Let us consider the stripe (see Fig.2)
\begin{equation}
{\cal {S}}:=\{(t,x)~|~t\in {\bf {R}}, ~x\in] 0 ,\pi /m [ \},
\end{equation}
with the metric tensor
\begin{equation}
g_{\mu\nu}(t, x)=\frac{1}{\sin^2 mx}
\left( \begin{array}{cr}
1&0\\0&-1 \end{array} \right).
\end{equation}
The coordinates $x$ and $t$ are associated with space ($x^1$)
and time ($x^0$) coordinates, respectively.
From (4.14) we get the positive constant curvature $R=2m^2$.
The isometry group of the manifold (4.13)-(4.14)
is the universal covering group of
${SL}(2.{\bf R})$.

The Liouville field $q=-\log\sin^2mx$ defines the traceless
energy-momentum tensor given by (4.5) and for the free field $\varphi$
we again get (4.7) (substituting $\theta_\pm$ by $x^\pm :=x\pm t$).
The difference is only in the domain of
admissible values of the coordinates $x^\pm$.
Since the time coordinate $t$ is not restricted,
the singularity lines are given by
\begin{equation}
x\pm t -\alpha_\pm =\frac{2\pi}{m}n_\pm ,
\end{equation}
where $n_\pm$ are any integers. In general, the corresponding
left and right `moving' lines intersect each other.
One can avoid intersection of singularities, assuming again
$\alpha_+=\alpha_-$.
In that case the free field $\varphi (t,x)$ is singular only on the zigzag
line (see Fig.2) and the class of singularities for the Cauchy data is
preserved.

\setcounter{equation}{0}

\section{Conclusion}

Analysis of 2d dilaton gravity shows
its similarity to the system of
massless particle in the corresponding gravitational field.
As it is shown in [8], the trajectories
of massless particle on the hyperboloid (4.1) are given by the generatrices
(4.8), and the `zigzag' line of Fig.2 is the trajectory on the stripe (4.13).

Since the dilaton (scalar) field $\phi= q-\varphi$ has
the same singularities as the free
field $\varphi$, we can associate the  dilaton field
with the massless scalar particles.
In the case of hyperboloid
the dilaton field describes two massless particles moving in `opposite'
directions (Fig.1), while for the stripe  there is only
one particle oscillating between the edges of `universe' (Fig.2).
By this correspondence we have the kinematical equivalence between the
`field theory' model and the particle system.
The dynamical equivalence can be achieved using the modified
energy-momentum tensor.
Adding the term $(g_{ab} \nabla^2 -\nabla_a\nabla_b)\exp {(1/2~\phi )}$
to the traceless tensor (4.5)  (see [12]),
we get the localized energy-momentum density,
which is concentrated at the  singularities of the configurations (4.7).

\vspace{3mm}

{\large\bf {Acknowledgments}}

One of the authors (G.J.) is indebted to Paul Sorba for the warm hospitality
at Annecy LAAP, where part of this work has been done. G.J. is also
grateful to W. Piechocki and G. Weigt for helpful discussions.

This work was supported by the grants from:
INTAS (96-0482), RFBR (96-01-00344) and
the Georgian Academy of Sciences.

%\end{document}

\vspace{0.3cm}
\setlength{\unitlength}{.1mm}

\begin{picture}(1500,800)

\put(0,400){\line(1,0){600}}
\put(250,0){\line(0,1){800}}

\multiput(50,50)(50,50){10}{\line(1,1){40}}

\multiput(50,750)(50,-50){11}{\line(1,-1){40}}

\put(200,740){$x^0$}
\put(550,360){$x^1$}
\put(400,400){\circle*{10}}
\put(480,700){$x^0>0$}
\put(360,200){$x^0<0$}

\put(900,400){\line(1,0){600}}
\put(950,0){\line(0,1){800}}
\put(1150,0){\line(0,1){800}}

\multiput(950,50)(50,50){4}{\line(1,1){40}}

\multiput(1150,250)(-50,50){4}{\line(-1,1){40}}

\multiput(950,450)(50,50){4}{\line(1,1){40}}

\multiput(1150,650)(-50,50){3}{\line(-1,1){40}}

\put(920,750){$t$}
\put(1450,375){$x$}
\put(1160,405){$2\pi/m$}
\put(910,405){$0$}
\put(950,400){\circle*{10}}
\put(1150,400){\circle*{10}}
\end{picture}

\hspace{1cm} Fig.1 $~(\Lambda <0)$ \hspace{7.0cm} Fig.2 $~(\Lambda >0)$

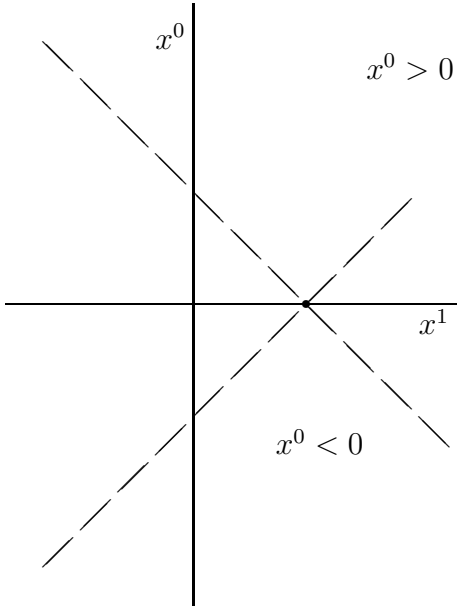
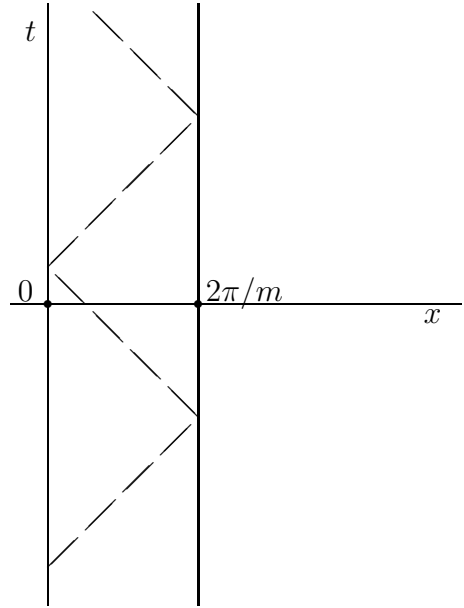
\end{document}